\documentclass[12pt]{iopart}
\usepackage[utf8]{inputenc}
\usepackage[T1]{fontenc}
\usepackage{xspace}
\expandafter\let\csname equation*\endcsname=\relax
\expandafter\let\csname endequation*\endcsname=\relax
\usepackage{amsmath,amssymb,amsfonts,amsthm}

\catcode`,\active

\catcode`\,12

\newcommand*\pFqskip{8mu}
\catcode`,\active
\newcommand*\pFq{\begingroup
        \catcode`\,\active
        \def ,{\mskip\pFqskip\relax}%
        \dopFq
}
\catcode`\,12
\def\dopFq#1#2#3#4#5{%
        {}_{#1}F_{#2}\biggl[\genfrac..{0pt}{}{#3}{#4};#5\biggr]%
        \endgroup
}

\renewcommand{\phi}{\varphi}
\newcommand{\pd}{\partial}
\newcommand{\ket}[1]{\ensuremath{|#1\rangle}\xspace}
\newcommand{\bra}[1]{\ensuremath{\langle #1|}\xspace}
\newcommand{\Braket}[3]{\ensuremath{\bra{#1}#2\ket{#3}}\xspace}
\newcommand{\braket}[2]{\ensuremath{\langle #1|#2\rangle}\xspace}
\newcommand{\under}[1]{_{#1}}

\begin{document}
\title[The Dunkl oscillator in the plane I]{The Dunkl oscillator in the plane I : superintegrability, separated wavefunctions and overlap coefficients}
\author{Vincent X. Genest}
\ead{genestvi@crm.umontreal.ca}
\address{Centre de recherches math\'ematiques, Universit\'e de Montr\'eal, C.P. 6128, Succursale Centre-ville, Montr\'eal, Qu\'ebec, Canada, H3C 3J7}
\author{Mourad E.H. Ismail}
\ead{mourad.eh.ismail@gmail.com}
\address{Department of Mathematics, University of Central Florida, Orlando, FL 32816, USA}
\address{Department of Mathematics, King Saud University, Riyadh, Saudi Arabia}
\author{Luc Vinet}
\ead{luc.vinet@umontreal.ca}
\address{Centre de recherches math\'ematiques, Universit\'e de Montr\'eal, C.P. 6128, Succursale Centre-ville, Montr\'eal, Qu\'ebec, Canada, H3C 3J7}
\author{Alexei Zhedanov}
\ead{zhedanov@yahoo.com}
\address{Donetsk Institute for Physics and Technology, Donetsk 83114, Ukraine}
\begin{abstract}
The isotropic Dunkl oscillator model in the plane is investigated. The model is defined by a Hamiltonian constructed from the combination of two independent parabosonic oscillators. The system is superintegrable and its symmetry generators are obtained by the Schwinger construction using parabosonic creation/annihilation operators. The algebra generated by the constants of motion, which we term the Schwinger-Dunkl algebra, is an extension of the Lie algebra $\mathfrak{u}(2)$ with involutions. The system admits separation of variables in both Cartesian and polar coordinates. The separated wavefunctions are respectively expressed in terms of generalized Hermite polynomials and products of Jacobi and Laguerre polynomials. Moreover, the so-called Jacobi-Dunkl polynomials appear as eigenfunctions of the symmetry operator responsible for the separation of variables in polar coordinates. The expansion coefficients between the Cartesian and polar bases (overlap coefficients) are given as linear combinations of dual $-1$ Hahn polynomials. The connection with the Clebsch-Gordan problem of the $sl_{-1}(2)$ algebra is explained. 
\end{abstract}
\section{Introduction}
This series of two papers is concerned with the analysis of the isotropic Dunkl oscillator model in the plane. The system will be shown to be superintegrable and the representations of its symmetry algebra will be related to different families of $-1$ orthogonal polynomials \cite{Genest-2012-1,Vinet-2012-1,Vinet-2012-2,Vinet-2011-2,Vinet-2011,Vinet-2012-4,Vinet-2012-3}.

A quantum system defined by a Hamiltonian $\mathcal{H}$ in $d$ dimensions is maximally \emph{superintegrable} if it admits $2d-1$ algebraically independent symmetry operators $S_i$, $1\leqslant i \leqslant2d-1$, that commute with the Hamiltonian 
\begin{equation*}
[\mathcal{H},S_i]=0,
\end{equation*}
where one of the operators is the Hamiltonian itself, e.g. $S_1\equiv\mathcal{H}$. For a superintegrable system described by a Hamiltonian of the form
\begin{equation*}
\mathcal{H}=\Delta+V(x),\quad \Delta=\frac{1}{\sqrt{g}}\sum_{ij}\pd_{x_i}(\sqrt{g}g^{ij})\pd_{x_j},
\end{equation*}
where $\Delta$ is the Laplace--Beltrami operator, the symmetries $S_i$ will be differential operators. In this case, the system is said to be superintegrable of order $\ell$ if   $\ell$ is the maximum order of the symmetry generators $S_i$ (other than $\mathcal{H}$). One of the most important features of superintegrable models is that they can be exactly solved.

When $\ell=1$, the constants of motion form a Lie algebra. When $\ell=2$, the symmetry algebra is quadratic \cite{Zhedanov-1992-2,Zhedanov-1992,Granovskii-1992,Vinet-1995,Zhedanov-1992-3}. Substantial work has been done on these systems which are now well understood and classified (see \cite{Daska-2001,Winter-2002,Kalnins-2005-I,Kalnins-2005-II,Kalnins-2005-III,Kalnins-2005-IV,Kalnins-2005-V,Miller-2001,Post-2012-2} and references therein). Further developments in the study of integrable systems include progress in the classification of superintegrable systems with higher order symmetry \cite{Kalnins-2011,TTW-2009,TTW-2010}, the examination of discrete/finite superintegrable models \cite{Post-2012} and the exploration of systems involving reflection operators \cite{Dunkl-2012,VDJ-2011,VDJ-2012,Nowak-2009,Plyu-1996,Plyu-1997,Post-2011,Rozen-1994,Rosler-1998,Ruffing-2000}.

We here examine the Dunkl oscillator in the plane. This model is possibly the simplest 2D system described by a Hamiltonian involving reflections and corresponds to the combination of two independent parabosonic oscillators. As will be shown, this system possesses many interesting properties. It is second-order superintegrable. Its symmetry algebra, which we term the Schwinger-Dunkl algebra, is obtained using parabosonic creation/annihilation operators in a way that parallels the Schwinger $\mathfrak{su}(2)$ realization in the case of the ordinary 2-dimensional isotropic oscillator; the Schwinger-Dunkl algebra is an extension of the Lie algebra $\mathfrak{u}(2)$ with involutions. The system admits separation of variables in both Cartesian and polar coordinates and its separated wavefunctions can be obtained explicitly in terms of the generalized Hermite, Jacobi and Laguerre polynomials. Furthermore, the study of this model and of the representations of its symmetry algebra will show remarkable occurrences of $-1$ orthogonal polynomials (OPs) families. The present paper is concerned with the exact solutions of the model, its superintegrability and the calculation of the overlap coefficients between the Cartesian and polar bases. The second paper of the series will focus on the representations of the symmetry algebra and the connections with $-1$ OPs.

Here is the outline of the paper. In Section 2, we define the Hamiltonian of the Dunkl oscillator and obtain its separated wavefunctions in Cartesian and polar coordinates. We also show that the symmetry operator responsible for the separation of variables in polar coordinates has the so-called Jacobi-Dunkl polynomials as eigenfunctions. In Section 3, we obtain the symmetry algebra of the model in terms of the parabosonic creation/annihilation operators. In Section 4, we show that the overlap coefficients between the Cartesian and polar bases are given by linear combinations of the dual $-1$ Hahn polynomials. In section 5, we exhibit the relationship between the Dunkl oscillator model and the Clebsch-Gordan problem of $sl_{-1}(2)$ \cite{Genest-2012-2,Zhedanov-2011}.

\section{The model and exact solutions}
The isotropic Dunkl oscillator model in the plane is defined by the Hamiltonian
\begin{equation}
\label{Hamiltonian}
\mathcal{H}=-\frac{1}{2}\left[(\mathcal{D}_{x}^{\mu_x})^2+(\mathcal{D}_{y}^{\mu_y})^2\right]+\frac{1}{2}\left[x^2+y^2\right],
\end{equation}
where the operator $\mathcal{D}_{x_i}^{\mu_i}$ is the Dunkl derivative
\begin{equation}
\label{Dunkl-D}
\mathcal{D}_{x_i}^{\mu_{x_i}}=\pd_{x_i}+\frac{\mu_{x_i}}{x_i}\left(\mathbb{I}-R_{x_i}\right),\quad x_i\in\{x,y\},
\end{equation}
with $\mathbb{I}$ the identity operator and $\pd_{x_i}=\frac{\pd}{\pd_{x_i}}$. The operator $R_{x_i}$ is the reflection operator with respect to the plane $x_i=0$. Hence the reflections in \eqref{Hamiltonian} have the action
\begin{equation*}
R_xf(x,y)=f(-x,y),\qquad R_yf(x,y)=f(x,-y).
\end{equation*}
In connection with the nomenclature of the standard harmonic oscillator, the model is called isotropic because the quadratic potential is $SO(2)$ invariant. For the full Hamiltonian \eqref{Hamiltonian} to have this symmetry requires of course $\mu_x=\mu_y$. Expanding the square of the Dunkl derivative, one finds
\begin{equation*}
(\mathcal{D}_{x_i}^{\mu_{x_i}})^2=\pd_{x_i}^2+2\,\frac{\mu_{x_i}}{x_i}\,\pd_{x_i}-\frac{\mu_{x_i}}{x_i^2}\left[\,\mathbb{I}-R_{x_i}\right].
\end{equation*}
The Schr\"odinger equation
\begin{equation}
\label{Schrodinger}
\mathcal{H}\Psi=\mathcal{E}\Psi,
\end{equation}
is manifestly separable in Cartesian coordinates. As shall be seen, even in the presence of reflections, \eqref{Schrodinger} also admits separation in polar coordinates. Separation of variables in more than one coordinate system is a signal of superintegrability. This occurs for the Dunkl oscillator because reflections can be viewed as rotations. We provide below the exact separated solutions of \eqref{Schrodinger}. Note that when $\mu_x=\mu_y=0$, the Hamiltonian \eqref{Hamiltonian} corresponds to the standard quantum Harmonic oscillator in the plane.
\subsection{Solutions in Cartesian coordinates}
Since the Hamiltonian \eqref{Hamiltonian} has the form 
\begin{equation*}
\mathcal{H}=\mathcal{H}_x+\mathcal{H}_y,
\end{equation*}
where $\mathcal{H}_x$ is the Hamiltonian of the one-dimensional Dunkl oscillator, it is obvious that the solutions to \eqref{Schrodinger} in Cartesian coordinates will be given by
\begin{equation*}
\Psi(x,y)=\psi(x)\psi(y),\qquad \mathcal{E}=\mathcal{E}_x+\mathcal{E}_y,
\end{equation*}
where $\psi(x_i)$ is an eigenfunction of the 1D Hamiltonian with energy eigenvalue $\mathcal{E}_{x_i}$. For the 1D oscillator $\mathcal{H}_x$, the Schr\"odinger equation reads
\begin{equation}
\label{1D-Eq}
\psi''(x)+\frac{2\mu_x}{x}\,\psi'(x)+(2\mathcal{E}_{x}-x^2)\psi(x)-\frac{\mu_x}{x^2}(\mathbb{I}-R_x)\psi(x)=0.
\end{equation}
Since $[\mathcal{H}_x,R_x]=0$, the eigenfunctions $\psi(x)$ may be chosen to have a definite parity $R_x\psi(x)=s_x \psi(x)$ with $s_x=\pm 1$. 

When $s_x=+1$, we have $R_x\psi^{+}(x)=\psi^{+}(x)$ and the equation \eqref{1D-Eq} has for (admissible) solutions
\begin{equation*}
\psi^{+}_{n}(x)= \sqrt{\frac{n!}{\Gamma(\textstyle{n+\mu_x+1/2)}}}\,e^{-x^2/2}\,L_{n}^{(\mu_x-1/2)}(x^2), 
\end{equation*}
where $L_{n}^{(\alpha)}(x)$ are the Laguerre polynomials \cite{Koekoek-2010} and where $\Gamma(x)$ denotes the gamma function. The eigenvalues are given by
\begin{equation*}
\mathcal{E}_x=2n+\mu_x+1/2,\qquad n\in\{0,1,2,\ldots\}.
\end{equation*}
When $s_x=-1$, we have $R_x\psi^{-}(x)=-\psi^{-}(x)$ and the solutions to \eqref{1D-Eq} are then
\begin{equation*}
\psi_{m}^{-}(x)=\sqrt{\frac{m!}{\Gamma(m+\mu_x+3/2)}}\,e^{-x^2/2}\,x\, L_{m}^{(\mu_x+1/2)}(x^2),
\end{equation*}
with eigenvalues 
\begin{equation*}
\mathcal{E}_x=2m+1+\mu_x+1/2,\qquad m\in \{0,1,2,\ldots\}.
\end{equation*}
From the orthogonality relation of the Laguerre polynomials \eqref{Laguerre}, it is easily seen that for $\mu_x>-1/2$, the eigenfunction $\psi_{n}^{\pm}$ obey
\begin{equation*}
\int_{-\infty}^{\infty}\psi_{n}^{s_x}(x)\,[\psi_{m}^{s_x'}]^{\ast}\,|x|^{2\mu_x}dx=\delta_{nm}\delta_{s_xs_x'},
\end{equation*}
where $x^{\ast}$ denotes complex conjugation. From the above considerations, it is clear that the eigenstates of $\mathcal{H}_x$ can be labeled by a single integer $n_x$ whose parity is that of the corresponding  wavefunction. For this purpose, we introduce the generalized Hermite polynomials \cite{Chihara-1978,Rozen-1994,Rosler-1998-2}
\begin{equation*}
H_{2n+p}^{\mu_x}(x)=(-1)^{n}\sqrt{\frac{n!}{\Gamma(n+p+\mu_x+1/2)}}\,x^{p}\,L_{n}^{(\mu_x-1/2+p)}(x^2),
\end{equation*}
where $p=0,1$. With this definition, the eigenfunctions of $\mathcal{H}_x$ can be expressed as
\begin{equation*}
\psi_{n_x}(x)=e^{-x^2/2}H_{n_x}^{\mu_x}(x),\qquad n_x\in\mathbb{N},
\end{equation*}
with energy eigenvalues $\mathcal{E}_{x}=n_x+\mu_x+1/2$. 

The eigenfunctions of the one-dimensional Dunkl oscillator are thus normalized and orthogonal on the weighted $L^{2}$ space endowed with the scalar product
\begin{equation}
\label{Ultra-2}
\braket{g}{f}=\int_{-\infty}^{\infty}g^{\ast}(x)f(x)\,|x|^{2\mu_x}\,dx.
\end{equation}
It is directly checked (see Appendix B) that the Dunkl derivative \eqref{Dunkl-D} is anti-Hermitian with respect to the scalar product \eqref{Ultra-2}. This establishes that the Hamiltonian \eqref{Hamiltonian} is Hermitian.

Using the above results for the one-dimensional Dunkl oscillator, it follows that the eigenstates of the full Hamiltonian \eqref{Hamiltonian} in the Cartesian basis satisfy
\begin{equation}
\label{Cartesian-Base}
\mathcal{H}\ket{n_x,n_y}=\mathcal{E}\ket{n_x,n_y},\qquad \mathcal{E}=n_x+n_y+\mu_x+\mu_y+1,
\end{equation}
where $n_x$, $n_y$ are non-negative integers. The wavefunctions have the expression
\begin{equation*}
\Psi_{n_x,n_y}(x,y)=e^{-(x^2+y^2)/2}H_{n_x}^{\mu_x}(x)H_{n_y}^{\mu_y}(y),
\end{equation*}
and they satisfy the orthonormality condition
\begin{equation*}
\int_{-\infty}^{\infty}\int_{-\infty}^{\infty}\Psi_{n_x,n_y}(x,y)\Psi^{\ast}_{n_x',n_y'}(x,y)|x|^{2\mu_x}|y|^{2\mu_y}dx\, dy=\delta_{n_xn_x'}\delta_{n_yn_y'},
\end{equation*}
provided that $\mu_x>-1/2$ and $\mu_y>-1/2$. For the 1D case see also \cite{Mukunda-1980,Rozen-1994}.
\subsection{Solutions in polar coordinates}
In the polar coordinate system
\begin{equation*}
x=\rho\cos\phi,\qquad y=\rho\sin\phi,
\end{equation*}
the Hamiltonian \eqref{Hamiltonian} can be written as
\begin{equation*}
\mathcal{H}=\mathcal{A}_{\rho}+\frac{1}{\rho^2}\mathcal{B}_{\phi},
\end{equation*}
where $\mathcal{A}_{\rho}$ has the expression
\begin{equation*}
\mathcal{A}_{\rho}=-\frac{1}{2}\left[\pd_{\rho}^2+\frac{1}{\rho}\,\pd_{\rho}\right]-\frac{1}{\rho}(\mu_x+\mu_y)\pd_{\rho}+\frac{1}{2}\rho^2,
\end{equation*}
and where $\mathcal{B}_{\phi}$ is given by
\begin{equation}
\label{B-Phi}
\mathcal{B}_{\phi}=-\frac{1}{2}\,\pd_{\phi}^2+(\mu_x\tan\phi-\mu_y\cot\phi)\pd_{\phi}+\frac{\mu_x\,(\mathbb{I}-R_x)}{2\cos^2\phi}+\frac{\mu_y\,(\mathbb{I}-R_y)}{2\sin^2\phi}.
\end{equation}
For separation of the Dunkl Laplacian in higher dimensions see \cite{DeBie-2011-2,DeBie-2011,Dunkl-2002}. The actions of the reflection operators are easily seen to be
\begin{equation*}
R_xf(\rho,\phi)=f(\rho,\pi-\phi),\quad R_y f(\rho,\phi)=f(\rho,-\phi).
\end{equation*}
Upon substitution of the separated wavefunction $\Psi(\rho,\phi)=P(\rho)\Phi(\phi)$ in \eqref{Schrodinger}, one obtains the pair of equations
\begin{subequations}
\begin{gather}
\label{radial}
\mathcal{A}_{\rho}P(\rho)-\mathcal{E}P(\rho)+\frac{m^2}{2\rho^2}P(\rho)=0,\\
\label{angular}
\mathcal{B}_{\phi}\Phi(\phi)-\frac{m^2}{2}\Phi(\phi)=0,
\end{gather}
\end{subequations}
where $m^2/2$ is the separation constant.

We start by examining the angular equation \eqref{angular}; it has the explicit form
\begin{equation}
\label{Angu}
\fl
\qquad
\Phi''-2(\mu_x\tan\phi-\mu_y\cot\phi)\Phi'-\frac{\mu_x(\mathbb{I}-R_x)}{\cos^2\phi}\Phi-\frac{\mu_y(\mathbb{I}-R_y)}{\sin^2\phi}\Phi+m^2\Phi=0.
\end{equation}
Since $[\mathcal{H},R_x]=[\mathcal{H},R_y]=0$, we shall label the eigenstates by the eigenvalues $s_x,s_y=\pm 1$ of the reflection operators $R_{x}$ and $R_{y}$. 

When $s_x=s_y=+1$, the equation \eqref{Angu} has the (admissible) solution
\begin{equation*}
\Phi^{++}_{n}=\textstyle
\sqrt{\frac{(2n+\mu_x+\mu_y)\Gamma(n+\mu_x+\mu_y)n!}{2\,\Gamma(n+\mu_x+1/2)\Gamma(n+\mu_y+1/2)}}
\displaystyle
P_{n}^{(\mu_x-1/2,\,\mu_y-1/2)}(x),
\end{equation*}
with $x=-\cos2\phi$ and where $P_{n}^{(\alpha,\beta)}(x)$ denotes the Jacobi polynomials \cite{Koekoek-2010}. This solution corresponds to the eigenvalue $m^2=4n(n+\mu_x+\mu_y)$ with $n\in\mathbb{N}$. 

When $s_x=s_y=-1$, the solutions reads
\begin{equation*}
\Phi^{--}_{n}=\textstyle
\sqrt{\frac{(2n+\mu_x+\mu_y)\Gamma(n+\mu_x+\mu_y+1)(n-1)!}{2\,\Gamma(n+\mu_x+1/2)\Gamma(n+\mu_y+1/2)}}
\displaystyle\, \sin\phi\cos\phi
\,P_{n-1}^{(\mu_x+1/2,\,\mu_y+1/2)}(x),
\end{equation*}
with variable $x=-\cos2\phi$ and eigenvalue $m^2=4n(n+\mu_x+\mu_y)$, $n\in\mathbb{N}$. It is understood that $P_{-1}^{(\alpha,\beta)}(x)=0$ and hence that $\Phi^{--}_{0}=0$.

When $s_x=+1$ and $s_y=-1$, the solution to equation \eqref{Angu} is given by
\begin{equation*}
\Phi^{+-}_{n}=
\textstyle
\sqrt{\frac{(2n+\mu_x+\mu_y)\Gamma(n+\mu_x+\mu_y+1/2)(n-1/2)!}{2\,\Gamma(n+\mu_x)\Gamma(n+\mu_y+1)}}
\displaystyle
\sin\phi\,P_{n-1/2}^{(\mu_x-1/2,\,\mu_y+1/2)}(x),
\end{equation*}
with variable $x=-\cos 2\phi$, eigenvalue  $m^2=4n(n+\mu_x+\mu_y)$ and where $n$ takes only positive half-integer values $n\in\{1/2,3/2,5/2,\ldots\}$.

Lastly, when $s_x=-1$ and $s_y=1$, the solution to the angular equation has the expression
\begin{equation*}
\Phi^{-+}_{n}=
\textstyle
\sqrt{\frac{(2n+\mu_x+\mu_y)\Gamma(n+\mu_x+\mu_y+1/2)(n-1/2)!}{2\,\Gamma(n+\mu_x+1)\Gamma(n+\mu_y)}}
\displaystyle
\cos\phi\,P_{n-1/2}^{(\mu_x+1/2,\,\mu_y-1/2)}(x),
\end{equation*}
with variable $x=-\cos 2\phi$, eigenvalue  $m^2=4n(n+\mu_x+\mu_y)$ and where $n$ takes only positive half-integer values $n\in\{1/2,3/2,5/2,\ldots\}$.

From the orthogonality relation of the Jacobi polynomials \eqref{Jacobi}, it is directly seen that the wavefunctions obey the orthogonality relation
\begin{equation*}
\int_{0}^{2\pi}\Phi^{s_x s_y}_{n}(\phi)\Phi^{s_x's_y'}_{m}(\phi)\,|\cos\phi|^{2\mu_x}|\sin\phi|^{2\mu_y}d\phi=\delta_{nm}\delta_{s_xs_x'}\delta_{s_ys_y'}.
\end{equation*}
As seen from the above considerations, the value of the separation constant is always $m^2=4n(n+\mu_x+\mu_y)$. When the product $s_xs_y=+1$ is positive, $n$ is a non-negative integer. When the product $s_xs_y=-1$ is negative, $n$ is a positive half-integer.

We now examine the radial equation \eqref{radial}. It reads
\begin{equation*}
P''(\rho)+\frac{1}{\rho}\,(1+2\mu_x+2\mu_y)P'(\rho)+\left(2\mathcal{E}-\rho^2-\frac{m^2}{\rho^2}\right)P(\rho)=0.
\end{equation*}
This equation has for solutions
\begin{equation*}
P_{k}(\rho)=
\textstyle{
\sqrt{\frac{2\,k!}{\Gamma(k+2n+\mu_x+\mu_y+1)}}\,
}\displaystyle
e^{-\rho^2/2}\rho^{2n}L_{k}^{(2n+\mu_x+\mu_y)}(\rho^2),
\end{equation*}
with the energy eigenvalues
\begin{equation*}
\mathcal{E}=2(k+n)+\mu_x+\mu_y+1,\qquad k\in \mathbb{N}.
\end{equation*}
Using the orthogonality relation of the Laguerre polynomials, one finds that the radial wavefunction obeys
\begin{equation*}
\int_{0}^{\infty}P_{k}(\rho)P_{k'}(\rho)\,\rho^{1+2\mu_x+2\mu_y}\,d\rho=\delta_{kk'}.
\end{equation*}
Hence the eigenstates of the Hamiltonian \eqref{Hamiltonian} in the polar basis can be denoted $\ket{k,n;s_x,s_y}$ and satisfy
\begin{equation}
\label{Polar-Base}
\mathcal{H}\ket{k,n;s_x,s_y}=\mathcal{E}\ket{k,n;s_x,s_y}, \quad \mathcal{E}=2(k+n)+\mu_x+\mu_y+1,
\end{equation}
where $k\in \mathbb{N}$ is a non-negative integer and where $n$ is a non-negative integer whenever the product $s_xs_y=+1$ is positive and a positive half-integer whenever the product $s_xs_y=-1$ is negative.

From the equations \eqref{Cartesian-Base} and \eqref{Polar-Base}, it is seen the states with a given energy $\mathcal{E}=N+\mu_x+\mu_y+1$ exhibit a $N+1$-fold degeneracy. Here are the first few eigenstates :
$$
\begin{array}{|c|c|c|}
\hline
\mathcal{E} & \ket{n_x,n_y} &\ket{k,n;s_x,s_y} \\
\hline
 \mathcal{E}_0=1+\mu_x+\mu_y& \scriptstyle\ket{0,0} & \scriptstyle{\ket{0,0;++}} \\
  \mathcal{E}_1=2+\mu_x+\mu_y&  \scriptstyle \ket{1,0},\,\ket{0,1}& \scriptstyle\ket{0,1/2;+-},\,\ket{0,1/2;-+} \\
 \mathcal{E}_2=3+\mu_x+\mu_y & \scriptstyle \ket{2,0},\,\ket{1,1},\,\ket{0,2} & \scriptstyle\ket{1,0;++},\,\ket{0,1;++},\,\ket{0,1;--} \\
 \mathcal{E}_3=4+\mu_x+\mu_y &\scriptstyle \ket{3,0},\,\ket{2,1},\,\ket{1,2},\,\ket{0,3}  & \scriptstyle \ket{1,1/2;+-},\ket{1,1/2;-+},\ket{0,3/2;+-},\ket{0,3/2;-+}\\
\mathcal{E}_4=5+\mu_x+\mu_y & \scriptstyle \ket{4,0}\,\ket{3,1},\,\ket{2,2},\,\ket{1,3},\,\ket{0,4} & \scriptstyle \ket{2,0;++},\,\ket{1,1;++},\,\ket{1,1;--},\,\ket{0,2;++},\,\ket{0,2;--}
\\
\hline
\end{array}
$$
The presence of these degeneracies can be attributed to the existence of a symmetry algebra that will be identified in Section 3.
\subsection{Separation of variables and Jacobi-Dunkl polynomials}
As is seen from \eqref{angular}, the separation of variables of the Schr\"odinger equation in polar coordinates is equivalent to the diagonalization of the operator $\mathcal{B}_{\phi}$. We thus have the following eigenvalue equation:
\begin{equation}
\label{Eigen}
\mathcal{B}_{\phi}\ket{k,n;s_x,s_y}=\frac{m^2}{2}\ket{k,n;s_x,s_y},\quad m^2=4n(n+\mu_x+\mu_y),
\end{equation}
where $n\in \mathbb{N}$ when $s_xs_y=1$ and $n\in \{1/2,3/2,\ldots\}$ when $s_xs_y=-1$. We shall consider the operator
\begin{equation*}
\mathcal{J}_2=i(x\mathcal{D}_{y}^{\mu_y}-y\mathcal{D}_{x}^{\mu_x}),
\end{equation*}
which in polar coordinates reads
\begin{equation*}
\mathcal{J}_2=i\left[\pd_{\phi}+\mu_y\cot\phi\,(\mathbb{I}-R_y)-\mu_x\tan\phi\,(\mathbb{I}-R_x)\right].
\end{equation*}
A simple computation shows that the square of the operator $\mathcal{J}_{2}$ is related to $\mathcal{B}_{\phi}$ in the following way:
\begin{equation}
\label{J2-B}
\mathcal{J}_2^2=2\mathcal{B}_{\phi}+2\mu_x\mu_y(\mathbb{I}-R_xR_y).
\end{equation}
Instead of the eigenvalue equation \eqref{Eigen}, we shall consider the one corresponding to the diagonalization of $\mathcal{J}_2$:
\begin{equation}
\label{Sqrt}
\mathcal{J}_{2}F_{\epsilon}(\phi)=\lambda_{\epsilon}F_{\epsilon}(\phi),
\end{equation}
where $\epsilon=s_xs_y=\pm 1$; this extra label on the eigenvalues $\lambda_{\epsilon}$ is allowed since $R_xR_y$ commutes with $\mathcal{J}_{2}$. It follows from \eqref{Eigen} and \eqref{J2-B} that the square of the eigenvalues $\lambda_{\epsilon}$ are given by
\begin{equation}
\label{Vlan-2}
\lambda_{+}^{2}=4n(n+\mu_x+\mu_y),\qquad \lambda_{-}^{2}=4(n+\mu_x)(n+\mu_y),
\end{equation}
where $n\in\mathbb{N}$ when $\epsilon=1$ and $n=\{1/2,3/2,\ldots\}$ when $\epsilon=-1$. Moreover, since $s_xs_y=\epsilon$, we have $R_x=\epsilon R_y$. To solve \eqref{Sqrt}, we consider the decomposition
\begin{equation}
\label{decompo}
F_{\epsilon}(\phi)=f_{\epsilon}^{+}(\phi)+f_{\epsilon}^{-}(\phi),
\end{equation}
where $R_yf_{\epsilon}^{\pm}(\phi)=\pm f_{\epsilon}^{\pm}(\phi)$. It is directly seen that given the decomposition \eqref{decompo}, the eigenvalue equation \eqref{Sqrt} is equivalent to the system of differential equations
\small
\begin{gather*}
\pd_{\phi}\left[f^{+}_{\epsilon}+f^{-}_{\epsilon}\right]+2\mu_y\cot\phi\,f^{-}_{\epsilon}-\mu_x\tan\phi\,[(1-\epsilon)f^{+}_{\epsilon}+(1+\epsilon)f^{-}_{\epsilon}]=-i\lambda_{\epsilon}[f^{+}_{\epsilon}+f^{-}_{\epsilon}],\\
\pd_{\phi}\left[-f^{+}_{\epsilon}+f^{-}_{\epsilon}\right]+2\mu_y\cot\phi\,f^{-}_{\epsilon}-\mu_x\tan\phi\,[(\epsilon-1)f^{+}_{\epsilon}+(1+\epsilon)f^{-}_{\epsilon}]=-i\lambda_{\epsilon}[f^{+}_{\epsilon}-f^{-}_{\epsilon}],
\end{gather*}
\normalsize
where the second equation was obtained from the first one by applying $R_y$. These equations are easily seen to be equivalent to
\begin{gather*}
\label{eq-1}
\pd_{\phi}f_{\epsilon}^{-}+2\mu_y\cot\phi f^{-}_{\epsilon}-\mu_x\tan\phi (1+\epsilon)f^{-}_{\epsilon}=-i\lambda_{\epsilon}f^{+}_{\epsilon},\\
\label{eq-2}
\pd_{\phi}f^{+}_{\epsilon}-\mu_x\tan\phi\,(1-\epsilon)f_{\epsilon}^{+}=-i\lambda_{\epsilon}f^{-}_{\epsilon}.
\end{gather*}
\subsubsection{The case $\epsilon=1$}\hfill

\noindent 
When $\epsilon=+1$, one has
\begin{subequations}
\begin{gather}
\label{vlan-2}
\pd_{\phi}f_{+}^{-}+2\mu_y\cot{\phi}f_{+}^{-}-2\mu_x\tan\phi f_{+}^{-}=-i\lambda_{+}f_{+}^{+}\\
\label{vlan}
\pd_{\phi}f_{+}^{+}=-i\lambda_{+}f_{+}^{-},
\end{gather}
\end{subequations}
Substituting \eqref{vlan} in \eqref{vlan-2} yields the equation
\begin{equation*}
\pd^{2}_{\phi}f_{+}^{+}+(2\mu_y\cot{\phi}-2\mu_x\tan\phi)\pd_{\phi}f_{+}^{+}+\lambda_{+}^2f_{+}^{+}=0.
\end{equation*}
Since $\lambda_{+}^2=4n(n+\mu_x+\mu_y)$, we directly obtain the result
\begin{equation*}
f_{+}^{+}=P_{n}^{(\mu_x-1/2,\,\mu_y-1/2)}(x),\quad f_{+}^{-}=\frac{i}{\lambda_{+}}\pd_{\phi}f_{+}^{+},
\end{equation*}
with $x=-\cos 2\phi$, eigenvalues $\lambda_{+}=\pm 2\sqrt{n(n+\mu_x+\mu_y)}$ and $n\in\mathbb{N}$. Consequently, for $\epsilon=+$, the eigensolutions of \eqref{J2-B} are given by
\begin{equation}
\label{solu}
F_{+}(\phi)=P_{n}^{(\alpha,\beta)}(x)+\frac{i}{\lambda_{+}}\pd_{\phi}P_{n}^{(\alpha,\beta)}(x),
\end{equation}
where the eigenvalues are given by
\begin{equation*}
\lambda_{+}=\pm 2 \sqrt{n(n+\mu_x+\mu_y)},\quad n\in \mathbb{N},
\end{equation*}
and where the parameters are $\alpha=\mu_x-1/2$, $\beta=\mu_y-1/2$ and $x=-\cos 2\phi$. When $\epsilon=+1$, $R_x=R_y$ and the operator $-i\mathcal{J}_2$ can be written as
\begin{equation*}
\Lambda_{\mu_x,\mu_y}=\pd_{\phi}+\frac{A'_{\mu_x,\mu_y}}{A_{\mu_x,\mu_y}}\frac{(\mathbb{I}-R_{y})}{2},
\end{equation*}
where
\begin{equation*}
A_{\mu_x,\mu_y}=2^{2(\mu_x+\mu_y)}(\sin |\phi|)^{2\mu_y}(\cos\phi)^{2\mu_x},
\end{equation*}
with $A'(\phi)=\pd_{\phi}A(\phi)$. This directly establishes that the polynomials defined by \eqref{solu} correspond to the so-called Jacobi-Dunkl polynomials studied in \cite{Chouchene-2005}.

It is possible to express the eigenfunctions of $\mathcal{J}_2$ in terms of the wavefunctions, which are eigenfunctions of $\mathcal{B}_{\phi}$. By taking the derivative of equation \eqref{Angu} with respect to $\phi$ for $s_x=s_y=1$  and adjusting the normalization, one obtains
\begin{equation*}
\pd_{\phi}\Phi^{++}_{n}(\phi)=2\sqrt{n(n+\mu_x+\mu_y)}\,\Phi^{--}_{n}(\phi)
.\end{equation*}
Upon substituting this result in \eqref{solu}, one finds that for $\epsilon=+1$, the eigenfunctions $F_{+}(\phi)$ of $\mathcal{J}_2$ and their corresponding eigenvalues are given by
\begin{equation}
\label{F-Plus}
F_{+}(\phi)=\Phi^{++}_{n}(\phi)\pm i\,\Phi^{--}_{n}(\phi),\quad \lambda_{+}=\pm 2\sqrt{n(n+\mu_x+\mu_y)}.
\end{equation}

\subsubsection{The $\epsilon=-1$ case}\hfill

\noindent
When $\epsilon=-1$, the equations \eqref{eq-1} and \eqref{eq-2} become
\begin{gather*}
\pd_{\phi}f_{-}^{-}+2\mu_y\cot\phi\,f_{-}^{-}=-i\lambda_{-}f_{-}^{+},\\
\pd_{\phi}f_{-}^{+}-2\mu_x\tan\phi\,f_{-}^{+}=-i\lambda_{-}f_{-}^{-}.
\end{gather*}
The first equation is easily rewritten as
\begin{align*}
\pd_{\phi}^2f_{-}^{+}+(2\mu_y\cot\phi-2\mu_x\tan\phi)\pd_{\phi}f_{-}^{+}+(\lambda_{-}^2-4\mu_x\mu_y)f_{-}^{+}-\frac{2\mu_x}{\cos^2\phi}f_{-}^{+}=0.
\end{align*}
Given the value of $\lambda_{-}^2$ defined in \eqref{Vlan-2}, we directly find
\begin{equation*}
f_{-}^{+}=\cos\phi\,P_{n-1/2}^{(\mu_x+1/2,\mu_y-1/2)}(x),\quad f_{-}^{-}=\frac{i}{\lambda_{-}}\left(\pd_{\phi}f_{-}^{+}-2\mu_x\tan\phi\, f_{-}^{+}\right)
\end{equation*}
with $x=-\cos 2\phi$. For $\epsilon=-1$, the eigenfunctions of $\mathcal{J}_2$ and their corresponding eigenvalues thus take the form
\begin{equation*}
F_{-}(\phi)=f_{-}^{+}(\phi)\pm f_{-}^{-}(\phi),\quad \lambda_{-}=\pm 2\sqrt{(n+\mu_x)(n+\mu_y)},
\end{equation*}
where $x=-\cos 2\phi$ and where $n$ is a positive half integer. In terms of the wavefunctions, a straightforward computation leads to the expression
\begin{equation}
\label{dompe-2}
F_{-}(\phi)=\Phi^{-+}_{n}(\phi)\mp i\,\Phi_{n}^{+-}(\phi),\quad \lambda_{-}=\pm 2\sqrt{(n+\mu_x)(n+\mu_y)}.
\end{equation}
Thus we have obtained the eigenfunctions of the operator $\mathcal{J}_{2}$ in terms of the wavefunctions, which are the eigenfunctions of $\mathcal{B}_{\phi}$.
\section{Superintegrability}
In this Section we show that the Dunkl oscillator model in the plane is superintegrable. We recover the spectrum of the Hamiltonian algebraically using the parabosonic creation/annihilation operators and obtain the symmetries using the Schwinger construction.

\subsection{Dynamical algebra and spectrum}
We first consider the dynamical algebra of the Dunkl oscillator model. We introduce two commuting sets of parabosonic creation/annihilation operators
\begin{align*}
A_{x_i}&=\frac{1}{\sqrt{2}}\left(x_i+\mathcal{D}_{x_i}^{\mu_{x_i}}\right),\qquad A_{x_i}^{\dagger}=\frac{1}{\sqrt{2}}(x_i-\mathcal{D}_{x_i}^{\mu_{x_i}}),
\end{align*}
where $x_i\in\{x,y\}$. These operators have the non-zero commutation relations
\begin{equation*}
[A_{x},A_{x}^{\dagger}]=\mathbb{I}+2\mu_x R_{x},\qquad [A_{y},A_{y}^{\dagger}]=\mathbb{I}+2\mu_yR_y.
\end{equation*}
In terms of creation/annihilation operators, the Hamiltonians $\mathcal{H}_{x}$ and $\mathcal{H}_y$ have the expression
\begin{equation}
\label{Rel-1}
\mathcal{H}_{x}=\frac{1}{2}\{A_{x},A_{x}^{\dagger}\},\qquad \mathcal{H}_y=\frac{1}{2}\{A_y,A_y^{\dagger}\},
\end{equation}
where $\{x,y\}=xy+yx$ denotes the anti-commutator. Thus the 2-dimensional Hamiltonian of the Dunkl oscillator \eqref{Hamiltonian} has the simple form
\begin{equation*}
\mathcal{H}=\frac{1}{2}\{A_{x},A_{x}^{\dagger}\}+\frac{1}{2}\{A_y,A_y^{\dagger}\}.
\end{equation*}
In the preceding Section, the eigenvalues $\mathcal{E}$ of $\mathcal{H}$ have been obtained analytically by solving the Schr\"odinger equation. They can also be obtained algebraically. Indeed, we  have the additional commutation relations
\begin{subequations}
\begin{gather}
\label{Rel-2}
[\mathcal{H}_{x_i},A_{x_i}]=-A_{x_i},\qquad [\mathcal{H}_{x_i},A_{x_i}^{\dagger}]=A_{x_i}^{\dagger}\\
\label{Rel-3}
\{A_{x_i},R_{x_i}\}=\{A_{x_i}^{\dagger},R_{x_i}\}=0,\qquad [\mathcal{H}_{x_i},R_{x_i}]=0,
\end{gather}
\end{subequations}
where $x_i\in\{x,y\}$. It is easily seen from the relations \eqref{Rel-1}, \eqref{Rel-2} and \eqref{Rel-3} that the operators $\mathcal{H}_{x_i}$, $A_{x_i}$, $A_{x_i}^{\dagger}$ and $R_{x_i}$ realize two independent copies of the parabosonic algebra which we have related to $sl_{-1}(2)$ in \cite{Zhedanov-2011}. It follows directly from the above commutation relations that
\begin{equation*}
\mathcal{E}_{x}=n_x+\mu_x+1/2,\quad \mathcal{E}_{y}=n_y+\mu_y+1/2, \quad n_x,n_y\in\mathbb{N}.
\end{equation*}
A direct computation shows that the action of the ladder operators $A_{x}$, $A_{x}^{\dagger}$ on the Cartesian eigenbasis $\ket{n_x,n_y}$ is given by
\begin{align}
\label{actions}
A_{x}^{\dagger}\ket{n_x,n_y}=\sqrt{[n_x+1]_{\mu_x}}\ket{n_x+1,n_y},\, A_{x}\ket{n_x,n_y}=\sqrt{[n_x]_{\mu_x}}\ket{n_x-1,n_y},
\end{align}
and that of the reflection $R_x$ by
\begin{equation*}
R_x\ket{n_x,n_y}=(-1)^{n_x}\ket{n_x,n_y},
\end{equation*}
where $[n]_{\mu}$ denotes the 'mu-numbers':
\begin{equation*}
[n]_{\mu}=n+\mu(1-(-1)^{n}).
\end{equation*}
Analogous formulas hold for the action of $A_{y}$, $A_{y}^{\dagger}$ and $R_{y}$.

As noted previously, the spectrum of the Hamiltonian $\mathcal{H}$ has the form
\begin{equation*}
\mathcal{E}_{N}=N+\mu_x+\mu_y+1,\quad N\in\mathbb{N},
\end{equation*}
and exhibits a $N+1$-fold 'accidental' degeneracy at level $N$. These degeneracies will be explained in terms of the irreducible representations of the symmetry algebra of the Dunkl oscillator.
\subsection{Superintegrability and the Schwinger-Dunkl algebra}
We now exhibit the symmetries of the Hamiltonian \eqref{Hamiltonian}. Let us consider the operator
\begin{equation*}
J_3=\frac{1}{4}\{A_{x},A_{x}^{\dagger}\}-\frac{1}{4}\{A_{y},A_{y}^{\dagger}\}=\frac{1}{2}\left(\mathcal{H}_x-\mathcal{H}_y\right).
\end{equation*}
It is clear that $[\mathcal{H},J_3]=0$ and that $J_3$ is the symmetry corresponding to separation of variables in Cartesian coordinates. Following the Schwinger construction \cite{Schwinger-1952}, we further introduce 
\begin{equation*}
J_2=\frac{1}{2i}\left(A_{x}^{\dagger}A_{y}-A_{x}A_y^{\dagger}\right).
\end{equation*}
A direct computation shows that $J_2$ is also a symmetry, i.e.  $[\mathcal{H},J_2]=0$. In addition, expressing the operator $J_2$ in terms of Dunkl derivatives shows that
\begin{equation*}
J_2=\frac{1}{2i}\big(x\mathcal{D}_{y}^{\mu_x}-y\mathcal{D}_{x}^{\mu_{x}}\big),
\end{equation*}
and hence $J_2=-\mathcal{J}_2/2$; it is thus seen from \eqref{J2-B} that $J_2$ is associated to the separation of variables in polar coordinates. To obtain the complete symmetry algebra, we define a third operator which also commutes with $\mathcal{H}$:
\begin{equation*}
J_1=\frac{1}{2}\left(A_{x}^{\dagger}A_{y}+A_{x}A_{y}^{\dagger}\right).
\end{equation*}
A direct computation show that the symmetry operators of the Dunkl oscillator in the plane satisfy the following algebra
\begin{gather*}
\{J_1,R_{x_i}\}=0,\qquad \{J_2,R_{x_i}\}=0,\qquad [J_3,R_{x_i}]=0,\\
[J_2,J_3]=iJ_1,\qquad [J_3,J_1]=i J_2,\\
[J_1,J_2]=i\left[J_3+J_3\,(\mu_xR_{x}+\mu_yR_{y})-\mathcal{H}(\mu_xR_{x}-\mu_{y}R_y)/2\right],
\end{gather*}
with $R_{x}^2=R_{y}^2=\mathbb{I}$, $x_i\in\{x,y\}$ and where the Hamiltonian $\mathcal{H}$ is a central element. We shall refer to the algebra generated by $J_1$, $J_2$, $J_3$, $R_x$, $R_y$ and $\mathcal{H}$ as the Schwinger-Dunkl algebra $sd(2)$; special cases of it have appeared in other contexts \cite{Genest-2012-2,VDJ-2011}. It is easily seen that $sd(2)$ is a deformation of the Lie algebra $\mathfrak{u}(2)$ by the two involutions $R_{x}$, $R_{y}$. The Schwinger-Dunkl algebra admits the Casimir operator \cite{Genest-2012-2} 
\begin{equation*}
C=J_1^2+J_2^2+J_3^2+\frac{1}{2}\mu_xR_x+\frac{1}{2}\mu_yR_y+\mu_x\mu_yR_xR_y,
\end{equation*}
which commutes with all the generators. A direct computation shows that in the present realization, the Casimir operator $C$ takes the value
\begin{equation*}
C=\frac{1}{4}\mathcal{H}^2-\frac{1}{4}.
\end{equation*}
Since $\mathcal{H}$ is a central element, we can define
\begin{equation*}
\widetilde{C}=C-\mathcal{H}^{2}/4+1/4,
\end{equation*}
and thus $\widetilde{C}=0$ in this realization.

The irreducible representations of the Schwinger-Dunkl algebra $sd(2)$ can be used to account for the degeneracies of the Hamiltonian \eqref{Hamiltonian}. We shall postpone this study for the second paper of the present series. Note that upon taking $\mu_x=\mu_y=0$ in the Schwinger-Dunkl algebra, the involutions cease to play an essential role and one recovers the well-known $\mathfrak{su}(2)$ symmetry algebra of the standard quantum harmonic oscillator in the plane.
\section{Overlap Coefficients}
In this section, we obtain the expansion (overlap) coefficients between the Cartesian and polar bases. These  expansion coefficients are denoted by $\braket{k,n;s_x,s_y}{n_x,n_y}$. It is clear that the coefficients will vanish unless the involved states $\ket{k,n;s_x,s_y}$ and $\ket{n_x,n_y}$ belong to the same energy eigenspace. The states in the polar basis are the eigenstates of the operator $\mathcal{B}_{\phi}$ given in \eqref{B-Phi} and satisfy
\begin{equation*}
B_{\phi}\ket{k;n;s_x,s_y}=\gamma_{n}\ket{k,n;s_x,s_y},\quad \gamma_{n}=2n(n+\mu_x+\mu_y),
\end{equation*}
with $n$ a non-negative integer whenever the product $s_xs_y=1$ and a positive half-integer otherwise. We can consider the relation
\begin{equation*}
\gamma_{n}\braket{k,n;s_x,s_y}{n_x,n_y}=\Braket{k,n;s_x,s_y}{\mathcal{B}_{\phi}}{n_x,n_y},
\end{equation*}
and expand the action of $\mathcal{B}_{\phi}$ on the Cartesian basis to obtain a recursion relation for the overlap coefficients. It will prove more convenient to investigate first the overlap coefficients between the Cartesian basis and the eigenbasis of a new operator $\mathcal{Q}$ related to $\mathcal{J}_2$. The eigenstates of this new operator $\mathcal{Q}$ will then be expanded in terms of the polar basis $\ket{k,n;s_x,s_y}$ to obtain the desired result. For this part, it is convenient to separate the two eigenvalue sectors corresponding to the value of the product $s_xs_y=\pm 1$.

\subsection{Overlap coefficients for $s_xs_y=+1$}
We start by expressing the energy eigenstates in the polar basis in terms of the eigenstates of $\mathcal{J}_2$. As is seen from \eqref{F-Plus}, the eigenvectors of the operator $\mathcal{J}_{2}$ with eigenvalues $\kappa_n^{\pm}$ that we denote $\ket{n,++}_{\mathcal{J}_2}$ and $\ket{n,+-}_{\mathcal{J}_2}$ are given by
\begin{align*}
&\ket{n,++}_{\mathcal{J}_2}=\frac{1}{\sqrt{2}}\Big(\ket{k;n;++}+i\ket{k,n;--}\Big),\quad \kappa_{n}^{+}=2\sqrt{n(n+\mu_x+\mu_y)},\\
&\ket{n,+-}_{\mathcal{J}_{2}}=\frac{1}{\sqrt{2}}\Big(\ket{k;n;++}-i\ket{k,n;--}\Big),\quad \kappa_{n}^{-}=-2\sqrt{n(n+\mu_x+\mu_y)},
\end{align*}
for $n\neq 0$. For $n=0$, one has
\begin{equation*}
\ket{0,++}_{\mathcal{J}_{2}}=\ket{k,0;++},\qquad \kappa_0^{+}=0.
\end{equation*}
We also recall that $R_{y}\ket{n,++}_{\mathcal{J}_2}=\ket{n,+-}_{\mathcal{J}_2}$. We now introduce the operator $\mathcal{Q}$ defined by
\begin{equation}
\label{Casi}
\mathcal{Q}=i\mathcal{J}_2R_{x}-\mu_xR_y-\mu_yR_x-(1/2)R_{x}R_{y}.
\end{equation}
The relevance of the operator $\mathcal{Q}$ will become clear in Section 5 when the connection between the Schwinger-Dunkl algebra and  the Clebsch-Gordan problem of $sl_{-1}(2)$ will be established. In the sector $s_xs_y=+1$, we have $R_x=R_y$ and $\mathcal{Q}$ may be written as
\begin{equation*}
\mathcal{Q}=i\mathcal{J}_{2}R_{y}-\mu_{x}R_{y}-\mu_yR_{y}-(1/2)\mathbb{I}.
\end{equation*}
For $n\neq0$, the eigenvalues $q_{n}^{\pm}$ and eigenvectors $\ket{n,+\pm}_{\mathcal{Q}}$ of $\mathcal{Q}$ are found to be
\begin{equation*}
\ket{n,++}_{\mathcal{Q}}=\frac{1}{\sqrt{2}}\left(\zeta_{n}\ket{n,++}_{\mathcal{J}_{2}}+\ket{n,+-}_{\mathcal{J}_2}\right),\; q_{n}^{+}=-2n-\mu_x-\mu_y-1/2,
\end{equation*}
and
\begin{equation*}
\ket{n,+-}_{\mathcal{Q}}=\frac{1}{\sqrt{2}}\left(-\zeta_{n}\ket{n,++}_{\mathcal{J}_2}+\ket{n,+-}_{\mathcal{J}_2}\right),\; q_{n}^{-}=2n+\mu_x+\mu_y-1/2,
\end{equation*}
where we have defined
\begin{equation*}
\zeta_{n}=\left[\frac{\mu_x+\mu_y-2i\sqrt{n(n+\mu_x+\mu_y)}}{2n+\mu_x+\mu_y}\right].
\end{equation*}
This amounts to the diagonalization of a $2\times 2$ matrix. We note that $\zeta_{n}\zeta^{\ast}_{n}=1$. When $n=0$, one has directly
\begin{equation*}
\ket{0,++}_{\mathcal{Q}}=\ket{0,++}_{\mathcal{J}_{2}},\quad q_{0}^{+}=-\mu_x-\mu_y-1/2.
\end{equation*}
It is possible to regroup the eigenvalues of $\mathcal{Q}$ into a single expression. We have
\begin{equation*}
q_{\ell}=(-1)^{\ell+1}(\ell+\mu_x+\mu_y+1/2),
\end{equation*}
and the eigenvectors are given by
\begin{align*}
\ket{q_{2j}}_{\mathcal{Q}}&=\frac{1}{\sqrt{2}}\Big(\zeta_{j}\,\ket{j,++}_{\mathcal{J}_{2}}+(1-\delta_{j0})\ket{j,+-}_{\mathcal{J}_2}\Big),\\
\ket{q_{2j+1}}_{\mathcal{Q}}&=\frac{1}{\sqrt{2}}\Big(-\zeta_{j+1}\,\ket{j+1,++}_{\mathcal{J}_2}+\ket{j+1,+-}_{\mathcal{J}_{2}}\Big).
\end{align*}
In the previous formulas, it should be understood that for the vector $\ket{q_0}$ the normalization factor $\sqrt{2}$ is not needed.

Having introduced the operator $\mathcal{Q}$, we examine the overlap coefficients between its eigenstates and the eigenstates of $\mathcal{H}$ in the Cartesian basis for a given energy level $\mathcal{E}_{N}$. In the sector $s_xs_y=+1$, the possible levels take the  energy values
\begin{equation*}
\mathcal{E}_{N}=N+\mu_x+\mu_y+1,
\end{equation*}
where $N$ is an even integer. The eigenspace $\mathcal{E}_{N}$ is spanned by the vectors
\begin{equation*}
\ket{0,N},\,\ket{1,N-1},\,\ldots,\ket{m,N-m},\,\ldots,\,\ket{N,0}.
\end{equation*}
We shall denote the overlap coefficients by
\begin{equation*}
\braket{q_{\ell}}{m,N-m}=M_{m,N}^{\ell}.
\end{equation*}
To obtain the expression for the expansion coefficients $M_{m,N}^{\ell}$, we start from the relation
\begin{equation}
\label{Dd}
q_{\ell}\,M_{m,N}^{\ell}=\Braket{q_{\ell}}{\mathcal{Q}}{m,N-m}.
\end{equation}
In terms of the parabosonic creation/annihilation operators, the operator $\mathcal{Q}$ reads
\begin{equation}
\label{D}
\mathcal{Q}=(A_{x}A_{y}^{\dagger}-A_{x}^{\dagger}A_{y})R_{x}-(\mu_x+\mu_y)R_{x}-(1/2)\mathbb{I}.
\end{equation}
Upon substituting \eqref{D} in \eqref{Dd} and using the actions \eqref{actions}, there comes
\begin{equation*}
q_{\ell}\,M_{m,N}^{\ell}=A_{m+1}\,M_{m+1,N}^{\ell}+B_{m}\,M_{m,N}^{\ell}+A_{m}\,M_{m-1,N}^{\ell},
\end{equation*}
where
\begin{align*}
A_{m}=(-1)^{m}\sqrt{[m]_{\mu_x}[N-m+1]_{\mu_y}},\quad B_m=(-1)^{m+1}(\mu_x+\mu_y)-1/2.
\end{align*}
It follows that the overlap coefficients $M_{m,N}^{\ell}$ can be expressed in terms of polynomials $\mathcal{P}_{m}(q_{\ell})$. Indeed, if we define
\begin{equation*}
M_{m,N}^{\ell}=M_{0,N}^{\ell} \,\mathcal{P}_{m}(q_{\ell}),
\end{equation*}
with $\mathcal{P}_{0}(q_{\ell})=1$, it transpires that $\mathcal{P}_{m}(q_{\ell})$ are polynomials of degree $m$ in the variable $q_{\ell}$ obeying the three-term recurrence relation
\begin{equation}
\label{recu}
q_{\ell}\,\mathcal{P}_{m}(q_{\ell})=A_{m+1} \mathcal{P}_{m+1}(q_{\ell})+B_{m} \mathcal{P}_{m}(q_{\ell})+A_{m} \mathcal{P}_{m-1}(q_{\ell}).
\end{equation}
Upon introducing the monic polynomials $\widehat{\mathcal{P}}_{m}(q_{\ell})$:
\begin{equation*}
\mathcal{P}_{m}(q_{\ell})=\frac{\widehat{\mathcal{P}}_{m}(q_{\ell})}{A_1\cdots A_{m}},
\end{equation*}
the recurrence relation \eqref{recu} becomes
\begin{equation}
\label{Recu-3}
q_{\ell}\,\widehat{\mathcal{P}}_{m}(q_{\ell})=\widehat{\mathcal{P}}_{m+1}(q_{\ell})+B_{m}\widehat{\mathcal{P}}_{m}(q_{\ell})+U_{m}\widehat{\mathcal{P}}_{m-1}(q_{\ell}),
\end{equation}
where 
\begin{equation}
\label{Coeff}
U_{n}=A_{n}^2=[m]_{\mu_{x}}[N-m+1]_{\mu_y}.
\end{equation}
Comparing the formulas \eqref{Recu-3} and \eqref{Coeff} with the formula \eqref{Recu-Coeff} of Appendix A, it is seen that the polynomials $\widehat{\mathcal{P}}_{m}(q_{\ell})$ correspond to the monic dual $-1$ Hahn polynomials $Q_{n}(x_{\ell};\alpha,\beta;N)$. We thus have
\begin{equation*}
\widehat{\mathcal{P}}_{m}(q_{\ell})=2^{-m}Q_{m}(x_{\ell},\alpha,\beta; N)
\end{equation*}
where the parameter identification is given by
\begin{equation*}
\alpha=2\mu_y+N+1,\quad \beta=2\mu_x+N+1,
\end{equation*}
and the variable $x_{\ell}$ takes the values
\begin{equation*}
x_{\ell}=(-1)^{\ell+1}(2\ell+2\mu_x+2\mu_y+1),\quad \ell=0,\ldots,N.
\end{equation*}
The value of $M_{0,N}^{\ell}$ can be obtained from the requirement that the overlap coefficients provide a unitary transformation between the two bases. Using the orthogonality relation \eqref{Ortho-Duaux} of the dual $-1$ Hahn polynomials, we obtain
\begin{equation*}
\braket{q_{\ell}}{m,N-m}=\sqrt{\frac{\omega_{N-\ell}}{U_{1}\cdots U_{m}}}\,Q_{m}(x_{\ell};\alpha,\beta;N),
\end{equation*}
where $\omega_{N-\ell}$ is the weight function \eqref{Poids-Duaux} of the dual $-1$ Hahn polynomials and $N$ is an even integer. It is seen from the formula \eqref{Poids-Duaux} of Appendix A that if $\mu_x>-1/2$ and $\mu_y>-1/2$, the weight function $\omega_{N-\ell}$ is positive for all $\ell\in\{0,\ldots,N\}$. The overlap coefficients obey the orthonormality relation
\begin{equation*}
\sum_{\ell=0}^{N}\braket{q_{\ell}}{m,N-m}\braket{n,N-n}{q_{\ell}}=\delta_{nm}.
\end{equation*}

It now possible to obtain the overlap coefficients between the Cartesian and polar wavefunctions of the Dunkl oscillator. We first observe that the eigenstates of $\mathcal{Q}$ have the expansion
\begin{equation*}
\ket{q_{2n+p}}=
\textstyle
\left[\frac{1+(-1)^{p}\zeta_{n+p}}{2}\right]
\displaystyle
\,\ket{k,n+p\,;++}+\,
\textstyle
\left[\frac{1-(-1)^{p}\zeta_{n+p}}{2i}\right]
\displaystyle
\ket{k,n+p\,;--},
\end{equation*}
with $p=0,1$; the formula is also valid for $n=p=0$. The inverse relations have the explicit form
\begin{align*}
\ket{k,n;++}=
\left[\frac{\zeta_n-1}{\zeta_{n}}\right]
\ket{q_{2n-1}}
-
\left[\frac{\zeta_n+1}{2i\zeta_{n}}\right]
\ket{q_{2n}},\\
\ket{k,n;--}=
\left[\frac{\zeta_n+1}{\zeta_{n}}\right]
\ket{q_{2n-1}}
+
\left[\frac{1-\zeta_n}{2i\zeta_{n}}\right]
\ket{q_{2n}},
\end{align*}
for $n\neq 0$. These formulas can be used directly to obtain the overlap coefficients
\begin{equation*}
\braket{k,n;++}{m,N-m},\quad \braket{k,n;--}{m,N-m},
\end{equation*}
as linear combinations of dual $-1$ Hahn polynomials. 
\subsection{Overlap coefficients for $s_xs_y=-1$}
The overlap coefficients in the parity sector $s_xs_y=-1$ are obtained similarly to the case $s_xs_y=1$. We again start by writing the energy eigenstates in the polar basis in terms of the eigenstates of the operator $\mathcal{J}_{2}$. It follows from the relation \eqref{dompe-2} that the eigenstates of the operator $\mathcal{J}_2$ with eigenvalues $\sigma_{n}^{\pm}$ that we denote by $\ket{n,-+}_{\mathcal{J}_2}$ and $\ket{n,--}_{\mathcal{J}_2}$ are given by
\begin{align*}
\ket{n,-+}_{\mathcal{J}_{2}}&=\frac{1}{\sqrt{2}}\Big(\ket{k,n;-+}-i\ket{k,n;+-}\Big),\quad \sigma_{n}^{+}=2\sqrt{(n+\mu_x)(n+\mu_y)},\\
\ket{n,--}_{\mathcal{J}_{2}}&=\frac{1}{\sqrt{2}}\Big(\ket{k,n;-+}+i\ket{k,n;+-}\Big),\quad \sigma_{n}^{-}=-2\sqrt{(n+\mu_x)(n+\mu_y)},
\end{align*}
where $n\in\{1/2,\,3/2,\ldots\}$. In this sector, the operator $\mathcal{Q}$ is equivalent to
\begin{equation*}
\mathcal{Q}=i\mathcal{J}_2R_{x}+(\mu_x-\mu_y)R_{x}+(1/2)\mathbb{Id}.
\end{equation*}
The eigenstates $\ket{n,-\pm}_{\mathcal{Q}}$ and eigenvalues $q_{n}^{\pm}$ of $\mathcal{Q}$ are easily found to be
\begin{equation*}
\ket{n,-+}_{\mathcal{Q}}=\frac{1}{\sqrt{2}}\left(\xi_{n}\ket{n,-+}_{\mathcal{J}_{2}}+\ket{n,--}_{\mathcal{J}_{2}}\right),\; q_{n}^{+}=1/2-2n-\mu_x-\mu_y,
\end{equation*}
 and
\begin{equation*}
\ket{n,--}_{\mathcal{Q}}=\frac{1}{\sqrt{2}}\left(-\xi_{n}\ket{n,-+}_{\mathcal{J}_{2}}+\ket{n,--}_{\mathcal{J}_{2}}\right),\;q_{n}^{-}=1/2+2n+\mu_x+\mu_y,
\end{equation*}
where
\begin{equation*}
\xi_{n}=\left[\frac{\mu_x-\mu_y+2i\sqrt{(n+\mu_x)(n+\mu_y)}}{2n+\mu_x+\mu_y}\right].
\end{equation*}
It is easily verified that $\xi_{n}\xi^{\ast}_{n}=1$. The eigenstates of $\mathcal{Q}$ can be grouped in a single expression. We write
\begin{equation*}
q_{\ell}=(-1)^{\ell+1}(\ell+\mu_x+\mu_y+1/2).
\end{equation*}
The eigenvectors have the expressions
\begin{gather*}
\ket{q_{2j+p}}_{\mathcal{Q}}=\frac{1}{\sqrt{2}}\Big((-1)^{p}\xi_{j+1/2}\,\ket{j+1/2,-+}_{\mathcal{J}_2}+\ket{j+1/2,--}_{\mathcal{J}_{2}}\Big).
\end{gather*}

We now compute the overlap coefficients between the eigenstates of $\mathcal{Q}$ and the eigenstates of $\mathcal{H}$ expressed in the Cartesian basis for a given energy level $\mathcal{E}_{N}$. In the sector $s_xs_y=-1$, the energy takes the values
\begin{equation*}
\mathcal{E}_{N}=N+\mu_x+\mu_y+1,
\end{equation*}
where $N$ is an odd integer. The eigenspace corresponding to $\mathcal{E}_{N}$ is spanned by the vectors
\begin{equation*}
\ket{0,N},\,\ket{1,N-1},\,\cdots \ket{m,N-m},\,\cdots,\,\ket{N,0}.
\end{equation*}
We denote the overlap coefficients by
\begin{equation*}
\braket{q_{\ell}}{m,N-m}=W_{m,N}^{\ell}.
\end{equation*}
The coefficients $W_{m,N}^{\ell}$ can be computed from the relation
\begin{equation}
\label{hyper}
q_{\ell}\,\braket{q_{\ell}}{m,N-m}=\Braket{q_{\ell}}{\mathcal{Q}}{m,N-m}.
\end{equation}
In terms of the parabosonic operators, the operator $\mathcal{Q}$ acting on the sector $s_{x}s_{y}$ reads
\begin{equation}
\label{Ultra}
\mathcal{Q}=(A_{x}A_{y}^{\dagger}-A_{x}^{\dagger}A_{y})R_{x}+(\mu_x-\mu_y)R_{x}+(1/2)\mathbb{Id}.
\end{equation}
Upon substituting \eqref{Ultra} in \eqref{hyper} and using the actions \eqref{actions}, one finds the recurrence relation
\begin{equation*}
q_{\ell}\,W_{m,N}^{\ell}=A_{m+1}\,W_{m+1,N}^{\ell}+\widetilde{B}_{m}\,W_{m,N}^{\ell}+A_{m}\,W_{m-1,N}^{\ell},
\end{equation*}
where 
\begin{equation*}
A_{m}=(-1)^{m}\sqrt{[m]_{\mu_x}[N-m+1]_{\mu_y}},\quad \widetilde{B}_{m}=(-1)^{m}(\mu_x-\mu_y)+1/2.
\end{equation*}
After writing
\begin{equation*}
W_{m,N}^{\ell}=W_{0,N}^{\ell}\,\mathcal{P}_{m}(q_{\ell}),
\end{equation*}
with $\mathcal{P}_{0}(q_{\ell})=1$ and introducing the monic polynomials
\begin{equation*}
\mathcal{P}_{m}(q_{\ell})=\frac{\widehat{\mathcal{P}}_{m}(q_{\ell})}{A_{1}\cdots A_{m}},
\end{equation*}
one finds that the polynomials $\widehat{\mathcal{P}}_{m}(q_{\ell})$ satisfy the three-term recurrence relation
\begin{equation}
\label{Recu-Dompe}
q_{\ell}\,\widehat{\mathcal{P}}_{m}(q_{\ell})=\widehat{\mathcal{P}}_{m+1}(q_{\ell})+\widetilde{B}_{m}\,\widehat{P}_{m}(q_{\ell})+\widetilde{U}_{n}\widehat{P}_{m-1}(q_{\ell}),
\end{equation}
with
\begin{equation*}
U_{m}=[m]_{\mu_x}[N-m+1]_{\mu_y}.
\end{equation*}
By comparing the recurrence relation \eqref{Recu-Dompe} with that of the dual $-1$ Hahn polynomials \eqref{Recu-Coeff}, one obtains
\begin{equation*}
\mathcal{P}_{m}(q_{\ell})=2^{-m}Q_{m}(x_{\ell},\alpha,\beta,N),
\end{equation*}
with the parameter identification
\begin{equation*}
\alpha=2\mu_x,\quad \beta=2\mu_y,
\end{equation*}
and the variable
\begin{equation*}
x_{\ell}=(-1)^{\ell}(2\ell+2\mu_x+2\mu_y+1).
\end{equation*}
The requirement that the overlap coefficients provide a unitary transformation leads to the relation
\begin{equation*}
\braket{q_{\ell}}{m,N-m}=\sqrt{\frac{w_{\ell}}{U_1\cdots U_m}}Q_{m}(x_{\ell},\alpha,\beta,N),
\end{equation*}
where $N$ is an odd integer. The overlap coefficients satisfy the orthogonality relation
\begin{equation*}
\sum_{\ell=0}^{N}\braket{q_{\ell}}{m,N-m}\braket{q_{\ell}}{n,N-n}=\delta_{nm}.
\end{equation*}

It is again possible to recover the overlap coefficients between the wavefunctions by expressing the eigenvectors of $\mathcal{Q}$ in terms of the eigenstates in the polar basis. One has
\begin{equation*}
\ket{q_{2j+p}}=
\textstyle
\left[\frac{1+(-1)^{p}\xi_{j+1/2}}{2}\right]
\displaystyle
\ket{k,j+1/2;-+}+
\textstyle
\left[\frac{(-1)^{p}\xi_{j+1/2}-1}{2i}\right]
\displaystyle
\ket{k,n;+-},
\end{equation*}
with $p\in\{0,1\}$. The inverse relation reads
\begin{gather*}
\ket{k,j+1/2;-+}=
\textstyle
\left[\frac{1+\xi_{j+1/2}}{2\xi_{j+1/2}}\right]
\displaystyle
\ket{q_{2j}}+
\textstyle
\left[\frac{\xi_{j+1/2}-1}{2i\xi_{j+1/2}}\right]
\displaystyle
\ket{q_{2j+1}},\\
\ket{k,j+1/2;+-}=
\textstyle
\left[\frac{\xi_{j+1/2}-1}{2\xi_{j+1/2}}\right]
\displaystyle
\ket{q_{2j}}+
\textstyle
\left[\frac{1+\xi_{j+1/2}}{2i\xi_{j+1/2}}\right]
\displaystyle
\ket{q_{2j+1}}.
\end{gather*}
Hence it is seen that the expansion coefficients between the Cartesian and polar bases are given in terms of linear combinations of dual $-1$ Hahn polynomials. These coefficients can also be expressed in integral form using the separated wavefunctions obtained in Section 2.
\section{The Schwinger-Dunkl algebra and the Clebsch-Gordan problem}
The Schwinger-Dunkl algebra and the dual $-1$ Hahn polynomials have both appeared in the examination of the Clebsch-Gordan problem for the Hopf algebra $sl_{-1}(2)$ \cite{Genest-2012-2,Zhedanov-2011}. In this Section, we explain the relationship between the two contexts. This will clarify the introduction of the operator $\mathcal{Q}$ in the previous Section.
\subsection{$sl_{-1}(2)$ Clebsch--Gordan coefficients and overlap coefficients}
The $sl_{-1}(2)$ algebra is generated by the elements $A_{0}$, $A_{\pm}$ and $R$ with the defining relations
\begin{align*}
[A_0,R]=0,\quad[A_0,A_{\pm}]=\pm A_{\pm},\quad \{A_{\pm},R\}=0,\quad \{A_{+},A_{-}\}=2A_{0},
\end{align*}
and $R^{2}=\mathbb{I}$. It admits the Casimir operator
\begin{align*}
\mathcal{Q}=A_{+}A_{-}R-A_0R+(1/2)R,
\end{align*}
which commutes will all the generators. This algebra has infinite-dimensional irreducible modules $V^{(\epsilon,\mu)}$ spanned by the basis vectors $v_{n}^{(\epsilon,\mu)}$, $n\in\mathbb{N}$. The action of the generators on the basis vectors is
\begin{align*}
&A_0v_{n}^{(\epsilon,\mu)}=(n+\mu+1/2)v_{n}^{(\epsilon,\mu)},\quad Rv_{n}^{(\epsilon,\mu)}=\epsilon (-1)^{n}v_{n}^{(\epsilon,\mu)},\\
&A_{+}v_{n}^{(\epsilon,\mu)}=\sqrt{[n+1]_{\mu}}v_{n+1}^{(\epsilon,\mu)},\quad A_{-}v_{n}^{(\epsilon,\mu)}=\sqrt{[n]_{\mu}}v_{n-1}^{(\epsilon,\mu)}.
\end{align*}
It is easily seen that $\mathcal{Q}\,v_{n}^{(\epsilon,\mu)}=-\epsilon\mu\,v_{n}^{(\epsilon,\mu)}$.

The $sl_{-1}(2)$ algebra is a Hopf algebra and has a non-trivial co-product. Upon taking the tensor product of two irreducible modules $V^{(\epsilon_1,\mu_1)}\otimes V^{(\epsilon_2,\mu_2)}$ spanned by the basis vectors $e_{n}^{(\epsilon_1,\mu_1)}\otimes e_{m}^{(\epsilon_2,\mu_2)}$, one obtains a third module $\widetilde{V}$ (in general not irreducible) by adjoining the action
\begin{gather*}
\widetilde{A}_0(v\otimes w)=(A_0v)\otimes w+v\otimes (A_0w),\quad \widetilde{R}(v\otimes w)=(Rv)\otimes (Rw),\\
\widetilde{A}_{\pm}(v\otimes w)=(A_{\pm}v)\otimes (Rw)+v\otimes(A_{\pm}w),
\end{gather*}
where $v\in V^{(\epsilon_1,\mu_1)}$ and $w\in V^{(\epsilon_2,\mu_2)}$. On $\widetilde{V}$, we have the Casimir element
\begin{equation*}
\widetilde{\mathcal{Q}}=(A_{-}^{(1)}A_{+}^{(2)}-A_{+}^{(1)}A_{-}^{(2)})R^{(1)}-(1/2)R^{(1)}R^{(2)}-\epsilon_1\mu_1R^{(2)}-\epsilon_{2}\mu_{2}R^{(1)},
\end{equation*}
where the superscripts indicate on which module the generators act; e.g. $A_{\pm}^{(1)}=A_{\pm}\otimes\mathbb{I}$.

The eigenvalues of $\widetilde{\mathcal{Q}}$ represent the irreducible modules $V^{(\epsilon_{i},\mu_{i})}$ appearing in the decomposition of $\widetilde{V}=\bigoplus_{i}V^{(\epsilon_i\mu_i)}$. The Clebsch-Gordan coefficients of $sl_{-1}(2)$ are the expansion coefficients between the direct product basis $e_{n}^{(\epsilon_1,\mu_1)}\otimes e_{m}^{(\epsilon_{2},\mu_2)}$ and the eigenbasis $f_{k}^{(\epsilon_1,\mu_i)}$ of the operator $\widetilde{\mathcal{Q}}$; this corresponds to the 'coupled' basis. Given the addition rule of $A_0$, one has 
\begin{align}
\label{RL}
f_{N}^{(\epsilon_1,\mu_i)}=\sum_{n_1+n_2=N}C^{\mu_1\mu_2\mu_i}_{n_1n_2N}\,e_{n_1}^{(\epsilon_1,\mu_1)}\otimes e_{n_2}^{(\epsilon_2,\mu_2)},
\end{align}
where $C^{\mu_1\mu_2\mu_i}_{n_1n_2N}$ are the Clebsch-Gordan coefficients, which were shown to be given in terms of dual $-1$ Hahn polynomials in \cite{Genest-2012-2,Zhedanov-2011}.

In our model, it is seen that the operators $\{\mathcal{H}_x,A_{x},A_{x}^{\dagger}\}$, and $\{\mathcal{H}_y,A_{y},A_{y}^{\dagger}\}$ realize the two $sl_{-1}(2)$ modules $V^{\mu_x}$ and $V^{\mu_y}$, with $\epsilon_{x}=\epsilon_{y}=1$. The Cartesian basis states $\ket{n_x,n_y}$ correspond to the direct product basis and the operator $\mathcal{Q}$ given in \eqref{Casi} corresponds to the Casimir operator $\widetilde{\mathcal{Q}}$. This explains the origin of the operator $\mathcal{Q}$ in our approach to the overlap coefficients.
\subsection{Occurrence of the Schwinger-Dunkl algebra}
In our model, the Schwinger-Dunkl algebra occurs as the symmetry algebra. The algebra $sd(2)$ also appears in the C.G. problem of $sl_{-1}(2)$ as a 'hidden' algebra. We illustrate how this comes about.

In the C.G. problem, it follows from \eqref{RL} that the following operators act as multiple of the identity:
\begin{align}
\label{Op}
A_{0}^{(1)}+A_0^{(2)},\quad\mathcal{Q}^{(1)},\quad \mathcal{Q}^{(2)},\quad R^{(1)}R^{(2)}.
\end{align}
In the direct product basis, in addition to the operators \eqref{Op}, the operators
\begin{align*}
\widehat{K}_0=(A_0^{(1)}-A_0^{(2)})/2,\quad \widehat{R}=R^{(1)}
\end{align*}
and $R^{(2)}$ are also diagonal. In the 'coupled' basis, in addition to \eqref{Op}, we have the Casimir operator $\widehat{K}_1=\widetilde{\mathcal{Q}}$ which is diagonal. Hence, the tensor product basis corresponds to having the operators \eqref{Op} plus $\widehat{K}_0$ and $\widehat{R}$ in diagonal form and the coupled basis corresponds to having the operators \eqref{Op} and $\widehat{K}_1$ in diagonal form. A direct computation shows that the set $\{\widehat{K}_0,\,\widehat{K}_1,\,\widehat{R}\}$ generates the Schwinger-Dunkl algebra \cite{Genest-2012-2}.

We have thus established the connection between our model and the Clebsch-Gordan problem of the algebra $sl_{-1}(2)$.

\section{Conclusion}
We considered the Dunkl oscillator model and showed that it is a superintegrable system. We have exhibited the symmetry algebra that we called the Schwinger-Dunkl algebra and we have obtained the exact solutions of the Schr\"odinger equation in terms of Jacobi, Laguerre and generalized Hermite polynomials in Cartesian and polar coordinates.  The expansion coefficients between the Cartesian and polar bases have been obtained exactly in terms of linear combinations of dual $-1$ Hahn polynomials and we established the connection between these overlap coefficients and the Clebsch-Gordan problem of the algebra $sl_{-1}(2)$.

The representations of the symmetry algebra of a superintegrable system explain how the degenerate eigenstates of this system are transformed into each other. In the second series of the paper, we shall consider the representations of the Schwinger-Dunkl algebra. As will be seen, these representations exhibit remarkable occurrences of other $-1$ polynomials.

It would be of interest to consider in a future study the 3D Dunkl oscillator model, which will provide another example of a superintegrable system with reflections. It was shown in \cite{Genest-2012-3} that the Bannai--Ito polynomials occur as Racah coefficients of the algebra $sl_{-1}(2)$. Given the connection between the 2D Dunkl oscillator and the Clecbsch-Gordan problem of $sl_{-1}(2)$, one can expect that the Bannai--Ito polynomials will occur in the description of the 3D Dunkl oscillator model.
\ack
V.X.G. holds a scholarship from \emph{Fonds de recherche du Qu\'ebec, Nature et technologies} (FRQNT). The research of M.E.H.I. is funded by NPST Program of King Saud University; project number 10-MAT1293-02 and DSFP of King Saud University in Riyadh. The research of L.V. is supported in part by the \emph{Natural Sciences and Engineering Research Council of Canada} (NSERC). L.V. and A.Z. wish to acknowledge the hospitality of the City University of Hong Kong where this work was initiated. A.Z. has further benefited from an extended stay at the \emph{Centre de Recherches Math\'ematiques} (CRM).
\appendix
\section*{Appendix A}
\setcounter{section}{1}
\subsection{Formulas for Laguerre polynomials}
The Laguerre polynomials $L_{n}^{(\alpha)}(x)$ are defined by \cite{Koekoek-2010}:
\begin{equation*}
L_{n}^{(\alpha)}(x)=\frac{(\alpha+1)_{n}}{n!}\,\pFq{1}{1}{-n}{\alpha+1}{x},
\end{equation*}
where $(a)_n=(a)(a+1)\cdots (a+n-1)$ is the Pochhammer symbol. They obey the orthogonality relation:
\begin{equation}
\label{Laguerre}
\int_{0}^{\infty}e^{-x}x^{\alpha}L_{m}^{(\alpha)}(x)L_{n}^{(\alpha)}(x)=\frac{\Gamma(n+\alpha+1)}{n!}\,\delta_{nm},
\end{equation}
for $\alpha>-1$.
\subsection{Formulas for Jacobi polynomials}
The Jacobi polynomials $P_{n}^{(\alpha,\beta)}(x)$ are defined by \cite{Koekoek-2010}:
\begin{equation*}
P_{n}^{(\alpha,\beta)}(x)=\frac{(\alpha+1)_{n}}{n!}\,\pFq{2}{1}{-n,n+\alpha+\beta+1}{\alpha+1}{\frac{1-x}{2}}
\end{equation*}
They obey the orthogonality relation:
\begin{equation}
\label{Jacobi}
\fl
\qquad
\int_{-1}^{1}(1-x)^{\alpha}(1+x)^{\beta}P_{m}^{(\alpha,\beta)}(x)P_{n}^{(\alpha,\beta)}(x)=
\textstyle
\frac{2^{\alpha+\beta+1}}{2n+\alpha+\beta+1}\frac{\Gamma(n+\alpha+1)\Gamma(n+\beta+1)}{\Gamma(n+\alpha+\beta+1)n !}
\displaystyle
\delta_{nm},
\end{equation}
provided that $\alpha>-1$ and $\beta>-1$.
\subsection{Formulas for dual $-1$ Hahn polynomials}
The monic dual $-1$ Hahn polynomials $Q_{n}(x;\alpha,\beta;N)$ have the recurrence relation \cite{Vinet-2012-1}:
\begin{equation*}
x\,Q_{n}(x)=Q_{n+1}(x)+b_{n}Q_{n}(x)+u_{n}Q_{n-1}(x).
\end{equation*}
The recurrence coefficients are given by:
\begin{equation}
\label{Recu-Coeff}
u_{n}=4[n]_{\xi}[N-n+1]_{\zeta}, \qquad b_{n}=
\begin{cases}
(-1)^{n+1}(2\xi+2\zeta)-1, & N\,\text{even},\\
(-1)^{n}(2\zeta-2\xi)-1, & N\,\text{odd},
\end{cases},\quad 
\end{equation}
where
\begin{equation*}
\xi=
\begin{cases}
\frac{\beta-N-1}{2}, & N\,\text{even},\\
\frac{\alpha}{2}, & N\,\text{odd},
\end{cases},\quad 
\zeta=
\begin{cases}
\frac{\alpha-N-1}{2} & N\,\text{even},\\
\frac{\beta}{2}, & N\,\text{odd}.
\end{cases}
\end{equation*}
They obey the orthogonality relation:
\begin{equation}
\label{Ortho-Duaux}
\sum_{\ell}^{N}\omega_{\ell}\,Q_{n}(x_{\ell})Q_{m}(x_\ell)=v_{n}\delta_{nm}
\end{equation}
The weight is given by
\begin{equation}
\label{Poids-Duaux}
\omega_{2j+q}=
\begin{cases}
\frac{(-1)^{j}(-m)_{j+q}}{j!}\frac{(1-\alpha/2)_{j}(1-\alpha/2-\beta/2)_{j}}{(1-\beta/2)_{j}(m+1-\alpha/2-\beta/2)_{j+q}}\frac{(1-\beta/2)_{m}}{(1-\alpha/2-\beta/2)_{m}}, & N\,\text{even},\\
\frac{(-1)^{j}(-m)_{j}}{j!}\frac{(1/2+\alpha/2)_{j+q}(1/2+\alpha/2+\beta/2)_{j}}{(1/2+\beta/2)_{j+q}(m+3/2+\alpha/2+\beta/2)_{j}}\frac{(1/2+\beta/2)_{m+1/2}}{(1+\alpha/2+\beta/2)_{m+1/2}}, & N\,\text{odd}.
\end{cases}
\end{equation}
where $q\in\{0,1\}$ and with $m=N/2$, $v_{n}=u_{1}\cdots u_{n}$. The grid points have the expression
\begin{equation*}
x_{\ell}=
\begin{cases}
(-1)^{\ell}(2\ell+1-\alpha-\beta), & N\,\text{even},\\
(-1)^{\ell}(2\ell+1+\alpha+\beta), & N\,\text{odd}.
\end{cases}
\end{equation*}
\section*{Appendix B}
\setcounter{section}{2}
We here indicate how it can be simply seen that the Dunkl derivative \eqref{Dunkl-D} is anti-Hermitian with respect to the scalar product \eqref{Ultra-2}. This is recorded for completeness and convenience. We need to check that
\small
\begin{align*}
\braket{\psi_2}{\mathcal{D}_x^{\mu}\psi_1}=\int_{-\infty}^{\infty}\psi_2^{*}(x)[\mathcal{D}_x^{\mu}\psi_{1}(x)]|x|^{2\mu_x}dx=-\int_{-\infty}^{\infty}[\mathcal{D}_x^{\mu}\psi_2(x)]^{*}\psi_{1}(x)|x|^{2\mu_x}dx=-\braket{\mathcal{D}_x^{\mu}\psi_2}{\psi_1},
\end{align*}
\normalsize
for all functions $\psi_1(x)$, $\psi_2(x)$ belonging to the $L^2$ space associated to the scalar produt \eqref{Ultra-2}. We split the computation in the four possible parity cases for $\psi_1(x)$, $\psi_2(x)$. This is sufficient since any function can be decomposed into its even and odd parts and since the scalar product \eqref{Ultra-2} is linear in its arguments.

In the even-even case, one has $\psi_{1}(x)=\psi_1(-x)$, $\psi_2(x)=\psi_2(-x)$. It follows that
\small
\begin{align*}
\braket{\psi_2}{\mathcal{D}_x^{\mu}\psi_1}=\int_{-\infty}^{\infty}\psi_{2}^{*}(x)[\mathcal{D}_{x}^{\mu}\psi_{1}(x)]|x|^{2\mu}dx=\int_{-\infty}^{\infty}\psi_2^{*}(x)[\pd_{x}\psi_{1}(x)]|x|^{2\mu}dx=0,
\end{align*}
\normalsize
since the integrand in odd. Similarly, we have $\braket{\mathcal{D}_x^{\mu}\psi_2}{\psi_1}=0$.

In the odd-odd case $\psi_{1}(x)=-\psi_1(-x)$, $\psi_2(x)=-\psi_2(-x)$ and one obtains
\small
\begin{align*}
\int_{-\infty}^{\infty}\psi_{2}^{*}(x)[\mathcal{D}_{x}^{\mu}\psi_{1}(x)]|x|^{2\mu}dx=\int_{-\infty}^{\infty}\psi_2^{*}(x)\left[\pd_{x}\psi_{1}(x)+\frac{2\mu}{x}\psi_1(x)\right]|x|^{2\mu}dx=0,
\end{align*}
\normalsize
since the integrand is odd. Similarly, we have $\braket{\mathcal{D}_x^{\mu}\psi_2}{\psi_1}=0$.

In the even-odd case, $\psi_1{(x)}=-\psi_{1}(-x)$, $\psi_{2}(x)=\psi_2(-x)$ and it follows that
\small
\begin{align*}
&\int_{-\infty}^{\infty}\psi_{2}^{*}(x)[\mathcal{D}_{x}^{\mu}\psi_{1}(x)]|x|^{2\mu}dx=\int_{-\infty}^{\infty}\psi_{2}^{*}(x)\left[\pd_{x}\psi_{1}(x)+\frac{2\mu}{x}\psi_1(x)\right]|x|^{2\mu}dx\\
&=2\int_{0}^{\infty}\psi_{2}^{*}(x)\left[\pd_{x}\psi_{1}(x)+\frac{2\mu}{x}\psi_1(x)\right]x^{2\mu}dx
\\
&=2\psi_1(x)\psi_2^{*}(x)x^{2\mu}\Big\rvert_{0}^{\infty}-2\int_{0}^{\infty}\pd_{x}[\psi_{2}^{*}(x)x^{2\mu}]\psi_{1}(x)dx+4\mu\int_{0}^{\infty}\psi_{2}^{*}(x)\psi_{1}(x)x^{2\mu-1}dx
\\
&=-2\int_{0}^{\infty}[\pd_{x}\psi_{2}^{*}(x)]\psi_{1}(x)|x|^{2\mu}dx=-\int_{-\infty}^{\infty}[\mathcal{D}_{x}^{\mu}\psi_{2}(x)]^{*}\psi_{1}(x)|x|^{2\mu}dx,
\end{align*}
\normalsize
where we have used the vanishing conditions on $\psi_1(x)$, $\psi_2(x)$ at infinity. It thus seen that
$
\braket{\psi_2}{\mathcal{D}_x^{\mu}\psi_1}=-\braket{\mathcal{D}_x^{\mu}\psi_2}{\psi_1}
$. 

In the even-odd case, one has $\psi_1{(x)}=\psi_{1}(-x)$, $\psi_{2}(x)=-\psi_2(-x)$ and one obtains
\small
\begin{gather*}
\int_{-\infty}^{\infty}\psi_{2}^{*}(x)[\mathcal{D}_{x}^{\mu}\psi_{1}(x)]|x|^{2\mu}dx=\int_{-\infty}^{\infty}\psi_{2}^{*}(x)[\pd_{x}\psi_{1}(x)]|x|^{2\mu}dx
=2\int_{0}^{\infty}\psi_2^{*}(x)[\pd_{x}\psi_{1}(x)]\,x^{2\mu}dx\\
=2\psi_{1}(x)\psi_2^{*}(x)x^{2\mu}\Big\rvert_0^{\infty}-2\int_{0}^{\infty}\left[\pd_{x}\psi_{2}^{*}(x)+\frac{2\mu}{x}\psi_{2}^{*}(x)\right]\psi_{1}(x)\,x^{2\mu}dx=-\int_{-\infty}^{\infty}[\mathcal{D}_{x}^{\mu}\psi_{2}(x)]^{*}\psi_1(x)|x|^{2\mu}dx,
\end{gather*}
\normalsize
where we have used the vanishing conditions on $\psi_1(x)$, $\psi_2(x)$ at infinity. Hence we have $
\braket{\psi_2}{\mathcal{D}_x^{\mu}\psi_1}=-\braket{\mathcal{D}_x^{\mu}\psi_2}{\psi_1}
$ and the result
\begin{align*}
\braket{\psi_2}{\mathcal{D}_x^{\mu}\psi_1}=-\braket{\mathcal{D}_x^{\mu}\psi_2}{\psi_1}.
\end{align*}
is established in all cases.

\section*{References}

\end{document}